\documentclass[]{JHEP3}

\usepackage{graphicx}
\usepackage{amsmath,amssymb,amscd,amsfonts,bm}

\title{Boosting Higgs discovery -- the forgotten channel}

\author{Christoph Hackstein\\
Institut f\"ur Theoretische Physik \\
Institut f\"ur Experimentelle Kernphysik \\
Karlsuhe Institute of Technology\\
76131 Karlsruhe, Germany\\
E-mail: \email{christoph.hackstein@kit.edu}
}
\author{Michael Spannowsky \\
Institute of Theoretical Science\\
University of Oregon\\
Eugene, OR  97403-5203, USA\\
E-mail: \email{mspannow@uoregon.edu}
}

\abstract{
Searches for a heavy Standard Model Higgs boson focus on the 'gold plated mode' where the Higgs decays to two leptonic $Z$ bosons. This channel provides a clean signature, in spite of the small leptonic branching ratios. We show that using fat jets the semi-leptonic $ZZ$ mode significantly increases the number of signal events with a similar statistical significance as the leptonic mode.}

\keywords{Higgs phenomenology}
\preprint{IEKP-KA/2010-18\\ KA-TP-27-2010}

\begin{document}


\section{Introduction}

The main task of the LHC is to understand electroweak symmetry breaking, {\it e.g.} by confirming or modifying the minimal Higgs mechanism of the Standard Model~\cite{Higgs:1964ia, Djouadi:2005gi}. The cleanest Higgs signatures arise from Higgs decays to gauge bosons, where the $Z$~\cite{Gunion:1985dj,Gunion:1987ke,Cahn:1986pe,goldplatedmode} or $W$~\cite{Baer:1998cm,Han:1998ma,Dittmar:1996ss} bosons decay leptonically. Seeing purely hadronic Higgs decays at the LHC is an attractive goal~\cite{Carena:2000yx,Mangano:2002wn,Drollinger:2001ym, cammin}, and recent developments in searches for boosted $H \to b\bar{b}$ decays are putting us into a promising position~\cite{Butterworth:2008iy,ATLASHW, Soper:2010xk,Plehn:2009rk}. Mixed leptonic and hadronic decay products of the Higgs boson appear for example in searches for $H \to \tau \tau$ in weak boson fusion (WBF) and have a similar reach as the purely leptonic mode~\cite{Rainwater:1999sd,rainwater2,Yildiz}. Most of the papers listed above focus on a low-mass Higgs boson, but the same question we can of course ask for any Higgs mass.

The reason for the overwhelming interest in a light Higgs boson is that 
global fits to electroweak precision measurements~\cite{lepelw} indicate that in the Standard Model the Higgs mass $m_H$ has to lie below $144~\text{GeV}$ at $95 \%$ CL. Direct searches at LEP exclude masses below 114.5 GeV~\cite{lepcol}, and CDF and DO report an exclusion of the 163-166~GeV mass window after collecting $2.1$-$5.4~ \text{fb}^{-1}$ of data~\cite{cdf:2009je}. In spite of this focus on light Higgs bosons we need to keep in mind that all of these measurements include theory assumptions, basically that there be no weak-scale modification of our Standard Model with its minimal Higgs sector. 

Once we allow for such modifications the Higgs boson might for example become heavier. While the lower bounds on the Higgs mass are set by experiment, upper bounds arise from theoretical considerations, including the tree-level unitarity requirements~\cite{Chanowitz:1984ne} and the triviality bound~\cite{Luscher:1988gc}. Requiring that the Higgs self coupling remains finite and conservatively assuming that the cut-off scale to where the Standard Model remains valid extends only to the Higgs boson mass itself, calculations on the lattice give an upper bound of $m_H<640~\text{GeV}$~\cite{Gockeler:1992zj}. Somewhere above this mass range the Higgs width will increase to a significant fraction of the Higgs mass, so we would not consider such a Higgs state fundamental. Such a broad Higgs resonance will generically become hard to observe as a well defined mass peak over backgrounds. In this paper we will focus on the mass range $300~\text{GeV} \le m_H \le 600~\text{GeV}$.

At the moment the experiments at the LHC are running and collecting data, but collisions at the $14~\text{TeV}$ design energy will not be possible within the next two or three years. The significant increase of the center-of-mass energy at the LHC compared to the Tevatron, before and after the upgrade to 14 TeV, will extend the Higgs boson exclusion and discovery reaches very rapidly. This is particularly obvious for the relatively easy intermediate mass range $m_H > 140~\text{GeV}$. The gluon-fusion channel~\cite{gluonfusion} yields the biggest production cross section for a Standard Model Higgs boson at the LHC. Once the branching ratio to $Z$ pairs becomes sizeable the so-called 'gold plated mode' $H \rightarrow ZZ^* \rightarrow 4 l$ results in a very clean final state and allows for a Higgs boson discovery up to $m_H= 600~\text{GeV}$ based on $10~\text{fb}^{-1}$ at $\sqrt{s}=14~\text{TeV}$~\cite{goldplatedmode}. In this mass region roughly $30$\% of the Higgs bosons decay to $Z$ bosons~\cite{hdecay}. The charged lepton mode can be complemented by $H \rightarrow ZZ^* \rightarrow l^+ l^- \nu \bar{\nu}$~\cite{Cahn:1986pe}.  Unfortunately, the fact that only $6\%$ of the $Z$ bosons decay to electrons or muons means that this gold plated mode is strongly statistics limited.
 Allowing for one of the $Z$ bosons to decay hadronically and hence including the $60$\% hadronic $Z$ decays increase the expected number of events. Note that 15\% of $Z$ bosons decay to $b \bar{b}$ pairs, but searching for this channel by imposing one or two additional $b$-tags contradicts our primary goal of increasing the number of signal events compared to the purely leptonic sample.\bigskip
 
 The semi-leptonic channel 
 \begin{equation}
 \label{eq:proc}
pp \rightarrow H \rightarrow ZZ \rightarrow (\ell \ell) \, (jj)
\end{equation}
 has not been given the attention it deserves in the context of heavy Higgs searches.
 This can be partly understood because it is very difficult to compete with the clean leptonic final state, and additional backgrounds like $Z+\text{jets}$ make the extraction of the semi-leptonic signal events a difficult task.
We argue that recent developments of subjet techniques~\cite{Butterworth:2008iy,Ellis:2009su,Ellis:2009me,Krohn:2009th} changes this assumption. If a heavy resonance ($H$) decays to intermediately resonances ($Z$) which subsequently decay to quarks, the final-state quarks will be highly collimated. Thus, the hadronic $Z$ decays can be collected in a `fat jet'. It has been shown for gauge bosons~\cite{fat_gauge}, Higgs bosons~\cite{Butterworth:2008iy,ATLASHW, Soper:2010xk,Plehn:2009rk,Kribs:2010hp,Falkowski:2010hi}, and top quarks~\cite{fat_top,Ellis:2009me,Plehn:2009rk} that we can achieve a successful QCD background rejection based on kinematic patterns of subjets inside the fat jet. The experimental signature in Eq.\,(\ref{eq:proc}) requires us to first reconstruct the boosted $Z$ boson using subjet techniques and then combine the hadronic $Z$ boson with a leptonic $Z$ decay to form a Higgs resonance.

\section{The gold plated mode} \label{sec:goldplated}

\begin{figure}[t]
\includegraphics[width=7.0cm]{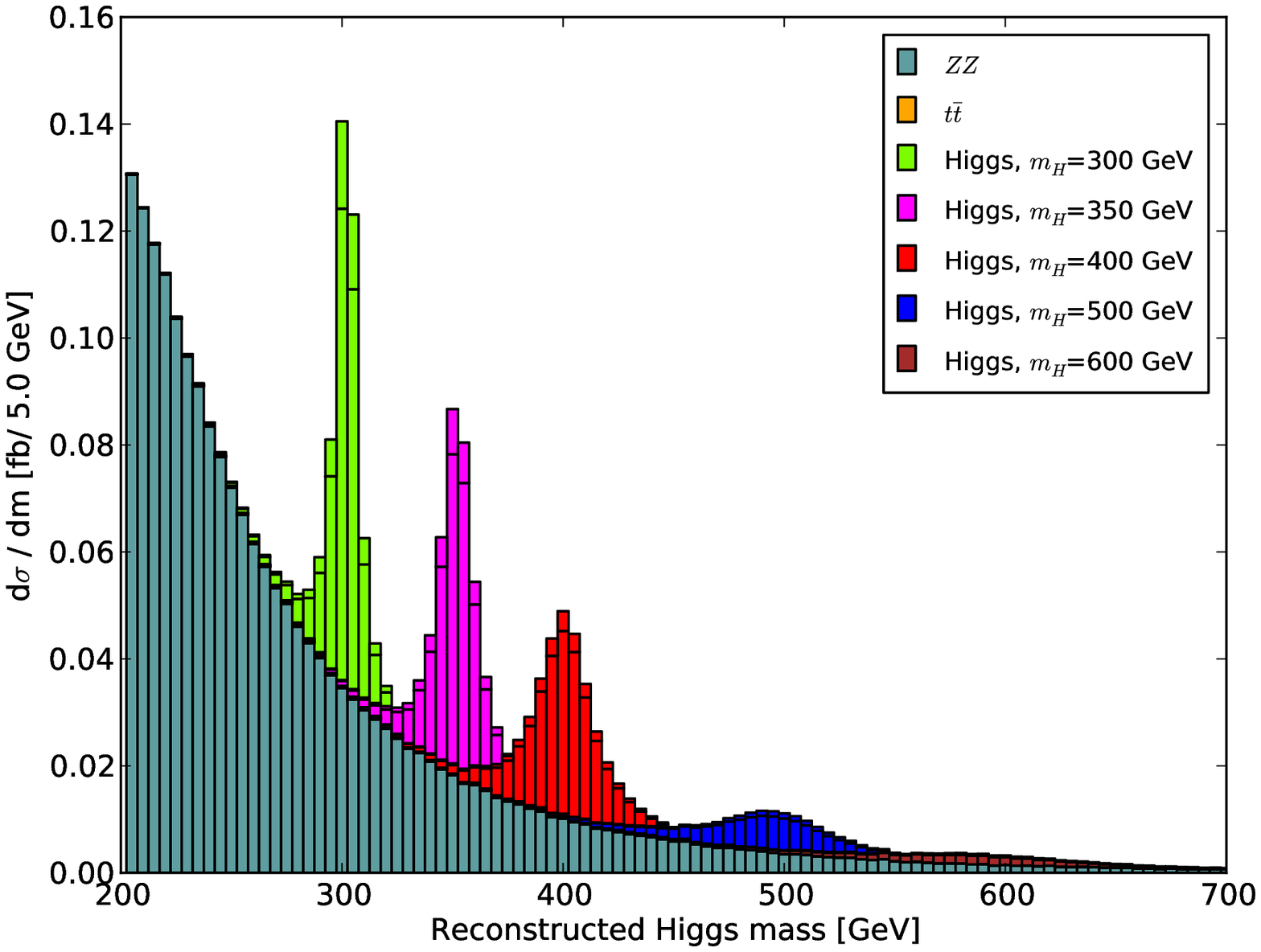}
\includegraphics[width=7.0cm]{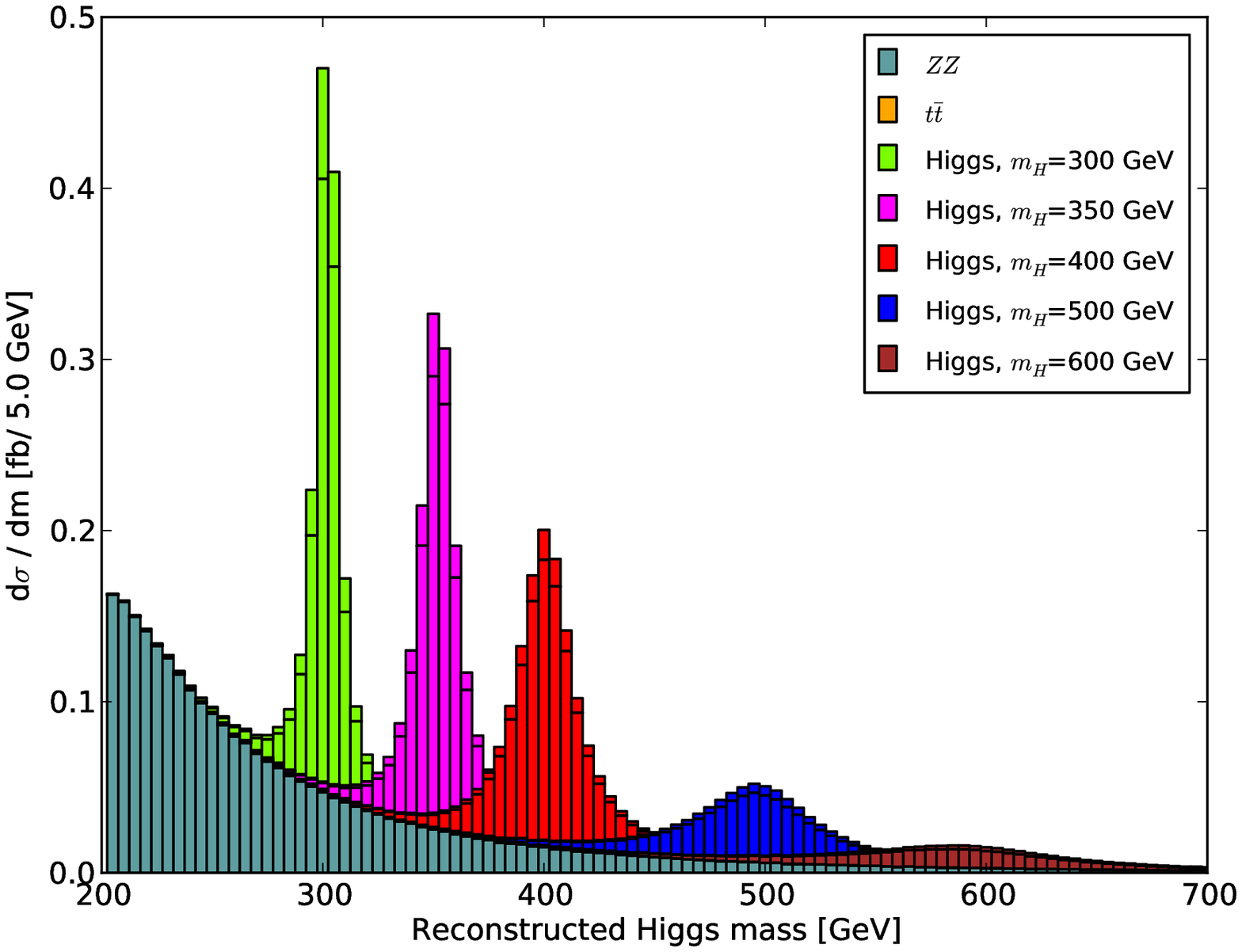}
\caption{Invariant mass distribution of the reconstructed $Z$ bosons in the semi-leptonic channel at 7 TeV (left) and 14 TeV (right) collider energy.}
\label{fig:lepresult}
\end{figure}

To be able to compare our semi-leptonic $ZZ$ chanel with the purely leptonic mode over the entire Higgs mass range we reproduce the results for the four-muon final state of~\cite{cmsnote} and find good agreement, see Figure~\ref{fig:lepresult}.

Throughout this paper we normalize the total rate for the 
gluon-fusion Higgs signal and the backgrounds to the next-to-leading order predictions. The NLO signal cross section we obtain by scaling the LO value from PYTHIA 6.4~\cite{Sjostrand:2006za} with a factor, $K =\sigma_\text{NLO}/\sigma_\text{LO}$ from HIGLU~\cite{higlu}. The transverse momentum distribution of the Higgs boson simulated with PYTHIA 6.4 approximates the full calculation with POWHEG very well~\cite{powheg}.  Weak boson fusion we include in the inclusive signal. The NLO corrections to this production process are known to be small~\cite{wbfnlo}, so within errors we assume K=1.0. The dominating background for the four-muon signature is continuum $ZZ$ production. We simulate this background using MadEvent~\cite{Alwall:2007st} and PYTHIA 6.4. Its NLO cross section comes from MCFM~\cite{mcfm}, giving us $7.39~\text{pb}$ at $7~\text{TeV}$ and $19.02~\text{pb}$ at $14~\text{TeV}$. 

To select a muon we demand it to be central
and sufficiently hard 
\begin{alignat}{5}
|y_\mu| < 2.5, \qquad \qquad \qquad 
p_{\text{T}, \mu} &> 7~\text{GeV} \qquad &&\text{for} \; |y_{\mu}| < 1.1 \notag \\
p_{\text{T}, \mu} &> 13~\text{GeV} &&\text{for} \; |y_{\mu}| > 1.1. 
\end{alignat}
The muons have to be isolated, that is the hadronic transverse energy in a cone of $R=0.3$ around the lepton has to be
$E_{\text{T}_{\text{hadronic}}} < 0.1~E_{\text{T}, \mu}$.
We accept events with at least four isolated muons passing the staggered $p_{\text{T}}$ cuts
\begin{equation}
p_{\text{T}, \mu}>15,15,12,8~\text{GeV} \; .
\end{equation}
The $Z$ bosons we reconstruct combining two oppositely charged isolated muons, requiring 
\begin{equation}
\label{eq:mz}
m_Z-10~\text{GeV} < m_{\mu \mu} < m_Z+10~\text{GeV}.
\end{equation}
\bigskip

For this analysis we consider five different Higgs-boson masses.
Because the Higgs width grows very fast with the Higgs mass~\cite{hdecay} we widen the mass windows for a reconstruction according to 
\begin{equation}
(300\pm30, \; 350\pm50, \; 400\pm50, \; 500\pm70, \; 600\pm100)~\text{GeV}.
\label{masswindows}
\end{equation}
The mass windows are completely dominated by the physical Higgs width. Detector effects like the lepton or jet energy scale will have only little effect, which means we keep the window for reconstructed Higgs mass for the leptonic and the semi-leptonic analyses.

\begin{table}[b]
\begin{tabular}{c||cc|cc||cc|cc}
 & \multicolumn{4}{c||}{7~TeV} & \multicolumn{4}{c}{14~TeV} \\
\hline
$ m_H$ [GeV] 
& $\sigma_S$ [fb] & $\sigma_B$ [fb] & $S/B$ & $S/\sqrt{B}_{10}$
& $\sigma_S$ [fb] & $\sigma_B$ [fb] & $S/B$ & $S/\sqrt{B}_{10}$ \\
\hline 
$300$ & 0.35 & 0.42 & 0.8 & 1.7 
      & 1.39 & 0.56 & 2.5 & 5.9 \\
$350$ & 0.35 & 0.38 & 0.9 & 1.8
      & 1.52 & 0.53 & 2.9 & 6.6  \\
$400$ & 0.28 & 0.21 & 1.3 & 1.9 
      & 1.34 & 0.31 & 4.4 & 7.6 \\
$500$ & 0.11 & 0.11 & 1.0 & 1.1 
      & 0.65 & 0.18 & 3.7 &  4.9  \\
$600$ & 0.05 & 0.07 & 0.7 & 0.6
      & 0.30  & 0.12 & 2.5 & 2.7  \\
\end{tabular}
\caption{Signal and background cross sections for the purely leptonic $H \to ZZ$ analysis.  The final significance we compute for $10~\text{fb}^{-1}$.}
\label{tab:reslep}
\end{table}

The purely leptonic channel is very clean and with four relatively hard muons not plagued by large background rates or large background uncertainties. Systematic errors should not be a problem since for $\sqrt{s}=14~\text{TeV}$ we find an outstanding signal-to-background ratio of $S/B > 1$ over the whole mass region. The results for collider energies of 7~TeV and 14~TeV we list in Table~\ref{tab:reslep}. The main distinguishing feature of signal and background is the four-muon invariant mass which we show in Figure~\ref{fig:lepresult}. Its signal shape is clearly distinguishable from the background, which makes this channel a save bet for a data driven side-bin analysis. While the significances shown for 7~TeV running will hardly give us an evidence for a heavy Higgs, at an energy of 14~TeV a discovery based on a modest integrated luminosity should not be a problem.

\section{The semi-leptonic channel} 
\label{sec:analysis}

For the semi-leptonic signature $pp \rightarrow H \rightarrow jj \ell \ell$ we need to consider $Z+\text{jets}$, $ZZ$, $t \bar{t}$ and $WZ$ backgrounds. 
For the main background $Z+\text{jets}$ the NLO rate after requiring $p_{\text{T}_{\text{jet}}} > 100~\text{GeV}$ is $33.91~(9.94)~\text{pb}$ for a collider energy of $14~(7)~\text{TeV}$~\cite{mcfm}. The NLO normalization of the $t\bar{t}$ rate is $875~(157.50)~\text{pb}$~\cite{Cacciari:2008zb} while for $WZ$ production we find  43.44 (17.31)~\cite{mcfm}. The $ZZ$ background corresponds to the numbers quoted in Section\,\ref{sec:goldplated}.

\begin{figure}[t]
\begin{center}
\includegraphics[width=10.0cm]{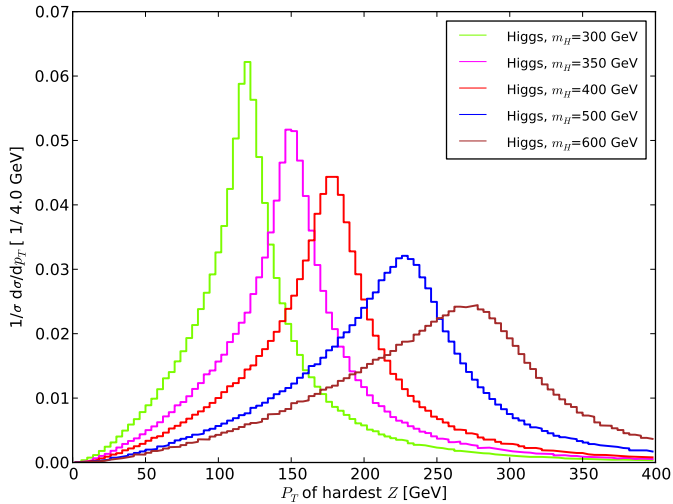}
\caption{$p_{\rm T}$ distribution of the leading $Z$ boson for different Higgs masses $m_H$, assuming $\sqrt{s}=14~\rm{TeV}$.}
\label{fig:ptZ}
\end{center}
\end{figure}

If a heavy Higgs boson decays to two $Z$ bosons the Higgs mass generates sizeable kinetic energy for the $Z$ bosons. In Figure\,\ref{fig:ptZ} we see that 70\% of the leading $Z$ bosons have $p_{\text{T}} > 150 ~\text{GeV}$ for $m_H=400~\text{GeV}$. The geometric distance between the $Z$ decay jets is roughly $\Delta R_{j_1,j_2} \simeq 2 m_Z /p_{\text{T}}$, which means that the inclusive Cambridge-Aachen (C-A) jet algorithm~\cite{Dokshitzer:1997in} with $R=1.2$ should be able to collect all $Z$ decay products in a fat jet.\bigskip 

Our analysis is based on grouping all final state particles after
showering and hadronization into detector cells of size $\Delta \eta
\times \Delta \phi = 0.1 \times 0.1$. This simulates the finite
resolution and thresholds of a calorimeter. We then combine all
particles in a cell and re-scale their total three-momentum such that
each cell has zero invariant mass. Only cells with energy above
$0.5~\text{GeV}$ we cluster into jets. The results of our analysis we
show for the following individual steps

\begin{description}
\item[Fat jet]
--- Our fat jet requirement on this calorimeter simulation uses the C-A algorithm implemented in FastJet~\cite{Cacciari:2005hq} with $R=1.2$. For this jet we require $|y_j| < 2$  and $p_{\text{T}_j} > 150~\text{GeV}$.

\item[Leptonic $Z$ reconstruction] 
--- As part of our selection cuts we ask for exactly two isolated muons with $p_\text{T}>15~\text{GeV}$ and $|\eta| < 2.5$. Their invariant mass has to match $m_Z \pm 10$~GeV. 

\item[Hadronic $Z$ reconstruction]
--- To reconstruct the hadronic $Z$ we follow Ref.~\cite{Butterworth:2008iy}. For the hardest jet in the event we undo the last stage of clustering. The two resulting subjets in the splitting $j \rightarrow j_1j_2$ are labeled such that $m_{j_1} > m_{j_2}$. If there is a significant mass drop, $m_{j_1} < \mu m_{j}$, and the splitting is not too asymmetric, $y= \Delta R_{j_1,j_2}^2 \text{min}(p_{T,j_1}^2, p_{T,j_2}^2) > y_\text{cut} m_j^2$, the jet $j$ is expected to be the resonance's neighborhood and the declustering stops, otherwise redefine $j$ to be equal to $j_1$. This process continues until the mass drop condition is met. If this does not happen the event is removed. We choose $\mu=0.67$ and $y_\text{cut}=0.09$. Varying $\mu = 0.33 - 0.67$ does not improve $S/\sqrt{B}$.
After the mass drop condition is met we filter the fat jet~\cite{Butterworth:2008iy}: the constituents of the two subjets which survive the mass drop condition are recombined with the higher resolution $R_\text{filt} =\text{min}(0.3,\Delta R_{j_1,j_2}/2)$ and the three hardest filtered subjets are required to give
$m_Z^\text{rec} = m_Z \pm 10~\text{GeV}$.

\item[Higgs reconstruction] 
--- If both $Z$ bosons in the signal are correctly reconstructed their invariant mass peaks around the Higgs boson mass,
$m_H^2 = (p_{Z,\text{lep}} + p_{Z,\text{had}})^2$.
The shape of the $m_H$ distribution is determined by the width of the Higgs boson and the ability of the algorithm to remove underlying event and initial state radiation from the hadronic $Z$ reconstruction. In practice, such an analysis would be combined with a likelihood fit or other elaborate statistical methods, taking into account systematic uncertainties. This is beyond the scope of this paper. Our choice of Higgs mass windows, Eq.(\ref{masswindows}), should give us a conservative estimate of the prospects of such an analysis.

\item[$ZZ$ separation]
--- After reconstructing the Higgs boson with a leptonic and a hadronic $Z$ boson $S/B$ can be further improved by requiring a maximum angular separation of $\Delta R_{ZZ} < 3.2$. For $Z+\text{jets}$ the angular separation of the reconstructed leptonic $Z$ and the fake-$Z$ from QCD jets often becomes large, to accomodate the large invariant (Higgs) mass. A similar effect we could achieve by scaling the $p_{\text{T}}$ cut on the hardest jet to higher values for larger Higgs masses. 

\item[pruning + trimming]
--- We know that a combination of pruning~\cite{Ellis:2009su,Ellis:2009me} and trimming~\cite{Krohn:2009th} helps discriminating the decay products of a color singlet resonance from QCD jets~\cite{Soper:2010xk}. All events passing the two $Z$ tags and the Higgs reconstruction we re-process using pruning and trimming on the massless cells of the event~\cite{Soper:2010xk}. For the pruning we use the C-A algorithm. For each pair of protojets to be combined we test if $\Delta R_{ij}> m_{\text{fat jet}}/p_{\text{T},{\text{fat jet}}}$ and $\min(p_{\text{T},i},p_{\text{T},j})/p_{\text{T},i+j}>0.1$. If both conditions hold true, the merging is vetoed and we discard the softer protojet. For the trimming we use the anti-$k_T$ algorithm~\cite{antiKT} to define the fat jet and the inclusive $k_T$ algorithm~\cite{KTjet} with a small cone $R=0.2$ for the subjet recombination. During trimming we keep all subjets with $p_{\text{T}_{\text{subjet}}} > 0.03~p_{\text{T}_{\text{fat jet}}}$.  Only if the pruned and trimmed masses of the leading jet are in the range $m_Z \pm 10~\text{GeV}$ we accept the event. 

\end{description}

\begin{table}[t]
\begin{center}
\begin{small}
\begin{tabular}{l||cc|cc|cc|cc}
$m_H$ [GeV]
& \multicolumn{2}{c|}{300} 
& \multicolumn{2}{c|}{400}
& \multicolumn{2}{c|}{500}
& \multicolumn{2}{c}{600} \\
\hline
$\sigma$ [fb]
&$\sigma_S$ & $\sigma_B$ 
&$\sigma_S$ & $\sigma_B$ 
&$\sigma_S$ & $\sigma_B$ 
&$\sigma_S$ & $\sigma_B$ \\
\hline \hline
selection 
& 3.37/0.89 & 907.3 
& 8.89/0.97 & 907.3
& 4.91/0.70 & 907.3
& 2.19/0.46 & 907.3 \\
$Z^\text{had}$ 
& 0.79/0.22 & 27.11 
& 3.81/0.42 & 27.11
& 2.36/0.35 & 27.11
& 1.11/0.25 & 27.11 \\
$m_H^\text{rec}$
& 0.46/0.17 & 1.02 
& 3.35/0.35 & 9.50
& 1.98/0.28 & 10.53
& 0.88/0.20 & 8.08 \\
$\Delta R_{ZZ}$  
& 0.45/0.17 & 1.00
& 2.99/0.35 & 7.93
& 1.52/0.28 & 6.52
& 0.60/0.15 & 3.82 \\
prun/trim
& 0.29/0.12 & 0.39 
& 2.02/0.24 & 3.97
& 1.11/0.18 & 3.33
& 0.46/0.12 & 1.97  \\
\hline
$S/B$ & 1.03 && 0.57 && 0.39 && 0.30 \\
$S/\sqrt{B}_{10}$ & 2.0 && 3.6 && 2.2 && 1.3 \\
\hline \hline
selection
& 17.97/3.83 & 6200
& 46.18/4.64 & 6200
& 29.48/3.87 & 6200 
& 15.08/2.90 & 6200 \\
$Z^\text{had}$ 
&  3.80/1.00 & 180.0
& 18.03/2.03 & 180.0 
& 13.49/1.98 & 180.0 
&  7.24/1.62 & 180.0 \\
$m_H^\text{rec}$
&  2.21/0.76 & 6.56
& 15.50/1.65 & 61.47
& 11.27/1.56 & 69.09 
&  5.75/1.24 & 54.16 \\
$\Delta R_{ZZ}$ 
&  2.18/0.76 &  6.45
& 13.94/1.55 & 52.22 
&  8.98/1.35 & 45.14
&  4.19/0.98 &  27.89 \\
prun/trim
& 1.34/0.48 & 2.10
& 8.96/1.07 & 19.21 
& 6.32/1.00 & 18.01 
& 3.15/0.77 & 11.83 \\
\hline
$S/B$ & 0.87 && 0.52 && 0.41 && 0.33 \\
$S/\sqrt{B}_{10}$ & 4.0 && 7.2 && 5.5 && 3.6

\end{tabular}
\end{small}
\caption{Signal and backgrounds for the
  semi-leptonic fat-jet analysis for a collider energy of 7~TeV (upper) and 14~TeV (lower). The expected significance
  is calculated for $10~\text{fb}^{-1}$. We show gluon fusion (left) and WBF (right) contributions separately for the signal cross sections. For the numbers of the expected significance we take both contributions into account.}
\label{tab:reshad}
\end{center}
\end{table}
 
 In Table~\ref{tab:reshad} we show the results for our LHC analysis for each of these steps. We separately give the gluon fusion and weak boson fusion signal rates and the background cross sections. After the reconstruction of the leptonic $Z$ and requiring  $p_{\text{T}}>150$~GeV for the leading jet the $Z+\text{jets}$ background still exceeds the signal by roughly a factor 1000. The hadronic $Z$ reconstruction in combination with the Higgs mass condition reduces this background tremendously and leaves us with typically $S/B \gtrsim 1/10$. Especially for a heavy Higgs boson the $\Delta R_{ZZ}$ cut proofs efficient against the $Z$+jets background. Finally, the combined pruning and trimming on the hadronic $Z$ improves $S/B$ over the whole considered Higgs mass range. The significance quoted can even further improved by including electrons in leptonic $Z$ reconstruction.

It is interesting to track the relative contributions of the gluon fusion and the weak boson fusion contributions to the inclusive signal. For small Higgs masses at 14~TeV collider energy the acceptance cuts leave us with a 80\%-20\% balance of the two channels. This enhancement as compared to the total rates is due to the generically larger Higgs transverse momentum in weak boson fusion, even if we do not cut on the tagging jets. For intermediate masses the weak boson fusion contribution drops to a 90\%-10\% ratio, because the Higgs transverse momentum of $p_{T,H} = \mathcal{O}(m_W)$ does not help to significantly boost the $Z$ decay products. Both channels are pushed far into their $p_{T,H}$ tails by the acceptance cuts. For large Higgs masses we know that the relative rate of weak boson fusion as compared to gluon fusion increases because of a logarithmic enhancement. This effect increases the relative weight of weak boson fusion back to 80\%-20\%. Of the different cuts only the hadronic $Z$ reconstruction shows a bias towards weak boson fusion, because the fat jet reconstruction is expected to benefit from the lower jet activity in this channel~\cite{Plehn:2009rk}. The final contribution from weak boson fusion ranges from $15\%$ for $400~\text{GeV} \leq m_H \leq 500~\text{GeV}$ to $30\%$ for either smaller or larger Higgs masses. This weak boson fusion contribution we expect to be a major handle for improving our results using advanced analysis methods. While here we do not make use of any of its kinematic features to suppress backgrounds an neural net could clearly include them.

\begin{figure}[t]
\includegraphics[width=7.0cm]{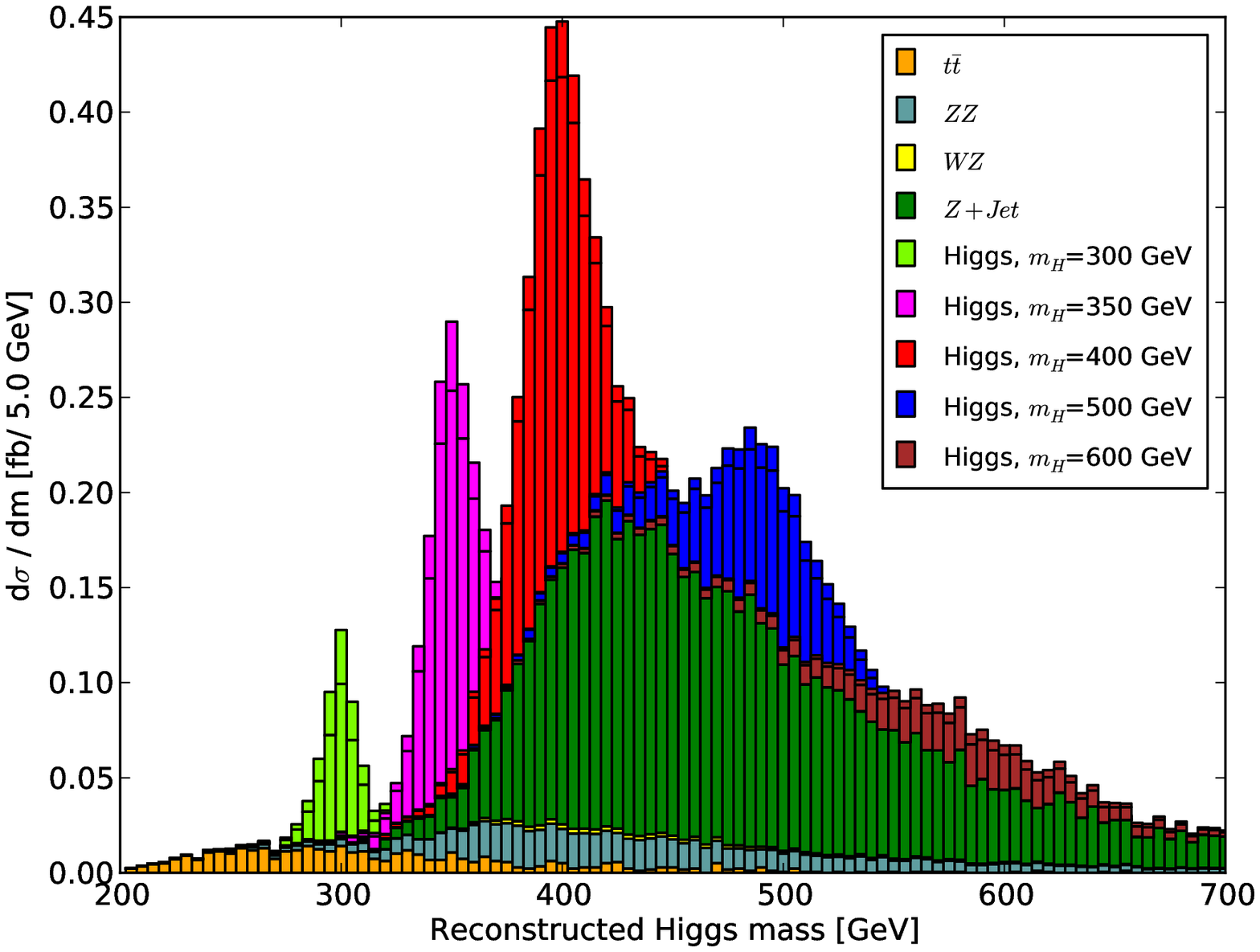}
\includegraphics[width=7.0cm]{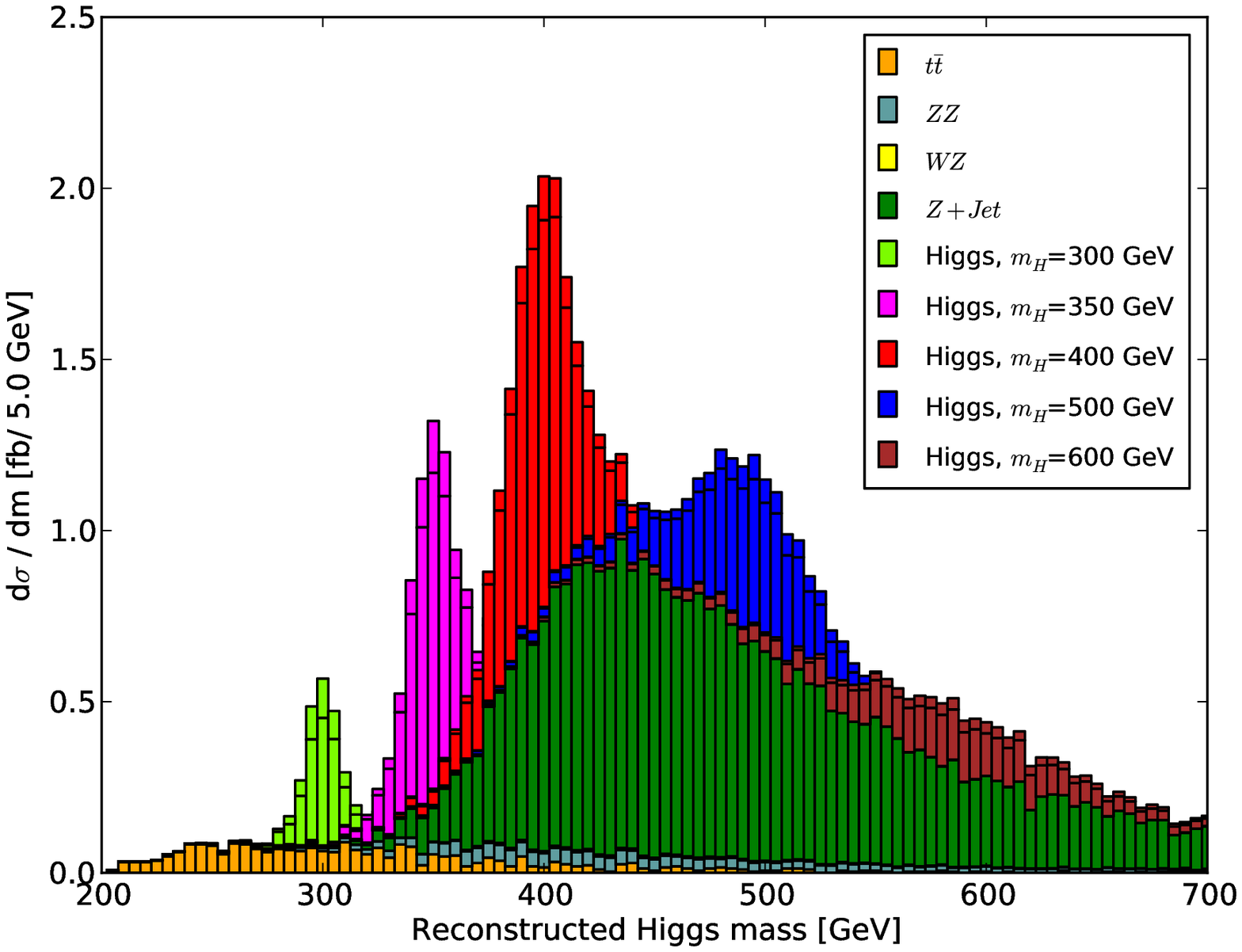}
\caption{Invariant mass distribution of the reconstructed $Z$ bosons, $m_{ZZ}$, in the semi-leptonic channel at $7$ TeV (left) and $14$ TeV (right) center-of-mass energy. A cut for the maximum angular separation of the $Z$ bosons has been applied, $\Delta R_{ Z_{\text{lep}},Z_{\text{had}} }$, as well as the combined usage of pruning and trimming. The signal consists of two parts. In every bin the top part shows the contribution of the WBF production process and the lower part the gluon fusion production process.
}
\label{fig:hadresult}
\end{figure}

In Figure~\ref{fig:hadresult} we show the reconstructed Higgs masses after pruning and trimming. The signal excess over backgrounds is clearly visible for $m_H = 300 - 500~\text{GeV}$.  By asking for boosted $Z$ bosons with a large angular separation we slightly shape the dominant $Z+\text{jets}$ background and generate a maximum around 450~GeV which should be taken into account in a side-bin analysis. \bigskip

As also shown in Table~\ref{tab:reshad} the cross sections for $\sqrt{s}=7~\text{TeV}$ are too small to allow for a Higgs discovery with early data. However, in new physics scenarios with a modified $ggH$ coupling this might change. A straightforward example is a chiral fourth generation for which electroweak precision data favors Higgs masses between $300$ and $500~\text{GeV}$~\cite{Kribs:2007nz}. Its loop contribution enhances the $ggH$ coupling by roughly a factor three. For an early LHC run at 7~TeV collecting $1~\text{fb}^{-1}$ integrated luminosity  our semi-hadronic analysis could then give 15 signal versus 4 background events for $m_H=400~\text{GeV}$.\bigskip

\section{Conclusions}
\label{sec:discussion}

We have shown that fat jet techniques will allow us to extract semi-leptonic $H \to ZZ$ decays at the LHC. To discriminate the signal from the large $Z+\text{jets}$ background we use a combination of mass drop searches and filtering based on large light-flavor C-A jets, as previously proposed to reconstruct a hadronic Higgs decay~\cite{Butterworth:2008iy}. 
For Higgs masses between 350~GeV and 500~GeV a successive reconstruction of the two $Z$ bosons and the Higgs boson extracts the inclusive signal at the $5 \sigma$ level based on $10~\text{fb}^{-1}$ at a 14~TeV LHC. Using additional information on the QCD structure of the event by employing a combined pruning/trimming analysis gives us typical signal-to-background ratios $S/B \sim 1/2$. We suggest that a thorough analysis of jet substructure techniques using early data in comparison to Monte-Carlo predictions is performed in the near future.

Comparing our results to the purely leptonic $ZZ$ channel at 14~TeV collider energy the leptonic signal on the one hand achieves $S/B>1$ while the semi-leptonic analysis only reaches $S/B\sim 0.33 - 0.87$. On the other hand, this is compensated by the larger number of signal events in the semihadronic channel. The semihadronic channel, which has never been considered to be a Higgs boson discovery channel before, can have as much statistical significance as the purely leptonic 'gold plated' mode. Heavy Higgs boson detection might greatly benefit from the orthogonal strength of our semi-leptonic $ZZ \to \ell \ell jj$ search, especially if for some reason the LHC should fall short of the full design energy or luminosity.

\acknowledgments
We thank Tilman Plehn and Manuel Zeise for valuable discussions and carefully reading the manuscript. This work was supported in part by the US Department of Energy under contract number DE-FG02-96ER40969. This research was supported in part by the Graduiertenkolleg ``High Energy Particle and Particle Astrophysics''.



\begin{thebibliography}{99}

\bibitem{Higgs:1964ia}
  P.~W.~Higgs,
  Phys.\ Lett.\  {\bf 12}, 132 (1964);
  P.~W.~Higgs,
  Phys.\ Rev.\ Lett.\  {\bf 13}, 508 (1964);
 F.~Englert and R.~Brout,
  Phys.\ Rev.\ Lett.\  {\bf 13}, 321 (1964).

\bibitem{Djouadi:2005gi}
  A.~Djouadi,
  Phys.\ Rept.\  {\bf 457}, 1 (2008);
 V.~Buescher and K.~Jakobs,
  Int.\ J.\ Mod.\ Phys.\  A {\bf 20}, 2523 (2005).


\bibitem{Gunion:1985dj}
  J.~F.~Gunion, P.~Kalyniak, M.~Soldate and P.~Galison,
  Phys.\ Rev.\  D {\bf 34}, 101 (1986).

 \bibitem{Gunion:1987ke}
  J.~F.~Gunion, G.~L.~Kane and J.~Wudka,
  Nucl.\ Phys.\  B {\bf 299}, 231 (1988).


\bibitem{Cahn:1986pe}
  R.~N.~Cahn and M.~S.~Chanowitz,
  Phys.\ Rev.\ Lett.\  {\bf 56}, 1327 (1986).


\bibitem{goldplatedmode}
  J.-C. Chollet et al., ATLAS note PHYS-NO-17 (1992); 
  L. Poggioli, ATLAS Note PHYS-NO-066 (1995); 
  D. Denegri, R. Kinnunen and G. Roullet, CMS-TN/93-101 (1993); 
  I. Iashvili R. Kinnunen, A. Nikitenko and D. Denegri, CMS TN/95-076; 
  D. Bomestar et al., Note CMS TN-1995/018; 
  C. Charlot, A. Nikitenko and I. Puljak, CMS TN/95-101; 
  G. Martinez, E. Gross, G. Mikenberg and L. Zivkovic, ATLAS Note ATL-PHYS-2003-001 (2003).




\bibitem{Baer:1998cm}
  H.~Baer and J.~D.~Wells,
  Phys.\ Rev.\  D {\bf 57}, 4446 (1998);
W.~Loinaz and J.~D.~Wells,
  Phys.\ Lett.\  B {\bf 445}, 178 (1998);
V. Cavasinni and D. Costanzo, ATL-PHYS-2000-013; 
K. Jakobs, ATLÐPHYSÐ2000Ð 008.
  
\bibitem{Han:1998ma}
  T.~Han and R.~J.~Zhang,
  Phys.\ Rev.\ Lett.\  {\bf 82}, 25 (1999);
    T.~Han, A.~S.~Turcot and R.~J.~Zhang,
  Phys.\ Rev.\  D {\bf 59}, 093001 (1999).
  
\bibitem{Dittmar:1996ss}
  M.~Dittmar and H.~K.~Dreiner,
  Phys.\ Rev.\  D {\bf 55}, 167 (1997);
K. Jakobs and Th. Trefzger, ATLAS Note 
ATLÐPHYSÐ2003Ð024.


\bibitem{Carena:2000yx}
  M.~S.~Carena {\it et al.}  [Higgs Working Group Collaboration],
  arXiv:hep-ph/0010338; 
   P.~C.~Bhat, R.~Gilmartin and H.~B.~Prosper,
  Phys.\ Rev.\  D {\bf 62}, 074022 (2000);
   K.~A.~Assamagan {\it et al.}  [Higgs Working Group Collaboration],
  arXiv:hep-ph/0406152.
  
\bibitem{Mangano:2002wn}
  M.~L.~Mangano, M.~Moretti, F.~Piccinini, R.~Pittau and A.~D.~Polosa,
  Phys.\ Lett.\  B {\bf 556}, 50 (2003).

\bibitem{Drollinger:2001ym}
  V.~Drollinger, T.~Muller and D.~Denegri,
  arXiv:hep-ph/0111312.

\bibitem{cammin}
J. Cammin and M. Schumacher, 
ATL-PHYS-2003-024[25] M. Bahr et al. , arXiv:0812.0529 [hep-ph]. 



\bibitem{Butterworth:2008iy}
  J.~M.~Butterworth, A.~R.~Davison, M.~Rubin and G.~P.~Salam,
  Phys.\ Rev.\ Lett.\  {\bf 100}, 242001 (2008).

\bibitem{ATLASHW}
  ATLAS Collaboration, "ATLAS Sensitivity to the Standard Model Higgs in the HW and HZ Channels at High Transverse Momenta",  ATL-PHYS-PUB-2009-088, ATL-COM-PHYS-2009-345.  

\bibitem{Soper:2010xk}
  D.~E.~Soper and M.~Spannowsky,
  arXiv:1005.0417.
  
  \bibitem{Plehn:2009rk}
  T.~Plehn, G.~P.~Salam and M.~Spannowsky,
  Phys.\ Rev.\ Lett.\  {\bf 104}, 111801 (2010).


\bibitem{Rainwater:1999sd}
T.~Plehn, D.~L.~Rainwater and D.~Zeppenfeld,
  Phys.\ Rev.\  D {\bf 61}, 093005 (2000);
  N.~Kauer, T.~Plehn, D.~L.~Rainwater and D.~Zeppenfeld,
  Phys.\ Lett.\  B {\bf 503}, 113 (2001).
M. Klute, ATLAS Note ATL-PHYS-2002-018 (2002); 
G. Azuelos and R. Mazini, ATLÐ PHYSÐ2003-004;
S. Asai et al. (ATLAS Collaboration), Eur. Phys. J. C32S2 (2004) 19.

\bibitem{rainwater2}
D.~L.~Rainwater and D.~Zeppenfeld,
  Phys.\ Rev.\  D {\bf 60}, 113004 (1999)
  [Erratum-ibid.\  D {\bf 61}, 099901 (2000)];

\bibitem{Yildiz}
H.D. Yildiz, M. Zeyrek and R. Kinnunen, CMSÐNOTEÐ2001/050;
K. Cranmer, B. Mellado, W. Quayle and S.L. Wu, Note ATLÐPHYSÐ2004-005


\bibitem{lepelw}
LEP Electroweak Working Group, http://lepewwg.web.cern.ch/LEPEWWG/;
  M.~W.~Grunewald,
  J.\ Phys.\ Conf.\ Ser.\  {\bf 110}, 042008 (2008).

\bibitem{lepcol}
The LEP Collaboratioin (ALEPH, DELPHI, L3 and OPAL), Phys. Lett. B565 (2003)

\bibitem{cdf:2009je}
    [CDF Collaboration and D0 Collaboration],
  arXiv:0911.3930 [hep-ex].
  
\bibitem{Chanowitz:1984ne}
  D.~A.~Dicus and V.~S.~Mathur,
  Phys.\ Rev.\  D {\bf 7}, 3111 (1973);
  M.~J.~G.~Veltman,
  Acta Phys.\ Polon.\  B {\bf 8}, 475 (1977);
  M.~S.~Chanowitz and M.~K.~Gaillard,
  Phys.\ Lett.\  B {\bf 142}, 85 (1984).

\bibitem{Luscher:1988gc}
  N.~Cabibbo, L.~Maiani, G.~Parisi and R.~Petronzio,
  Nucl.\ Phys.\  B {\bf 158}, 295 (1979);
    R.~F.~Dashen and H.~Neuberger,
  Phys.\ Rev.\ Lett.\  {\bf 50}, 1897 (1983);
  D.~J.~E.~Callaway,
  Nucl.\ Phys.\  B {\bf 233}, 189 (1984);
  M.~Luscher and P.~Weisz,
  Phys.\ Lett.\  B {\bf 212}, 472 (1988);
   M.~Luscher and P.~Weisz,
  Nucl.\ Phys.\  B {\bf 318}, 705 (1989).

\bibitem{Gockeler:1992zj}
  A. Hasenfratz, Quantum Fields on the Computer, Ed. M. Creutz, World ScientiÞc, 
  Singapore, 1992, p. 125;
  M.~Gockeler, H.~A.~Kastrup, T.~Neuhaus and F.~Zimmermann,
  Nucl.\ Phys.\  B {\bf 404}, 517 (1993).

\bibitem{gluonfusion}  
  H.~M.~Georgi, S.~L.~Glashow, M.~E.~Machacek and D.~V.~Nanopoulos,
  Phys.\ Rev.\ Lett.\  {\bf 40}, 692 (1978).

\bibitem{hdecay}  
    A.~Djouadi, J.~Kalinowski and M.~Spira,
  Comput.\ Phys.\ Commun.\  {\bf 108}, 56 (1998).

\bibitem{Ellis:2009su}
  S.~D.~Ellis, C.~K.~Vermilion and J.~R.~Walsh,
  Phys.\ Rev.\  D {\bf 80}, 051501 (2009).
  
\bibitem{Ellis:2009me}
  S.~D.~Ellis, C.~K.~Vermilion and J.~R.~Walsh,
  arXiv:0912.0033 [hep-ph].

\bibitem{Krohn:2009th}
  D.~Krohn, J.~Thaler and L.~T.~Wang,
  JHEP {\bf 1002}, 084 (2010).

\bibitem{Kribs:2010hp}
  G.~D.~Kribs, A.~Martin, T.~S.~Roy and M.~Spannowsky,
  Phys.\ Rev.\  D {\bf 81}, 111501 (2010);
  G.~D.~Kribs, A.~Martin, T.~S.~Roy and M.~Spannowsky,
  arXiv:1006.1656 [hep-ph].

\bibitem{Falkowski:2010hi}
  C.~R.~Chen, M.~M.~Nojiri and W.~Sreethawong,
  arXiv:1006.1151 [hep-ph];
  A.~Falkowski, D.~Krohn, J.~Shelton, A.~Thalapillil and L.~T.~Wang,
  arXiv:1006.1650 [hep-ph].

\bibitem{fat_gauge}
 J.~M.~Butterworth, J.~R.~Ellis and A.~R.~Raklev,
JHEP {\bf 0705}, 033 (2007);	
T.~Han, D.~Krohn, L.~T.~Wang and W.~Zhu,
 JHEP {\bf 1003}, 082 (2010);
 ATLAS Collaboration CERN-OPEN-2008-020.

\bibitem{fat_top}
 G.~Brooijmans,
  ATL-PHYS-CONF-2008-008 and ATL-COM-PHYS-2008-001, Feb.~2008
 J.~Thaler and L.~T.~Wang,
  JHEP {\bf 0807}, 092 (2008);
   D.~E.~Kaplan, K.~Rehermann, M.~D.~Schwartz and B.~Tweedie,
  Phys.\ Rev.\ Lett.\  {\bf 101}, 142001 (2008);
 L.~G.~Almeida, S.~J.~Lee, G.~Perez, G.~Sterman, I.~Sung and J.~Virzi,
  Phys.\ Rev.\  D {\bf 79}, 074017 (2009);
 L.~G.~Almeida, S.~J.~Lee, G.~Perez, I.~Sung and J.~Virzi,
  Phys.\ Rev.\  D {\bf 79}, 074012 (2009);
  T.~Plehn, M.~Spannowsky, M.~Takeuchi and D.~Zerwas,
  arXiv:1006.2833 [hep-ph];
    S.~Chekanov and J.~Proudfoot,
  Phys.\ Rev.\  D {\bf 81}, 114038 (2010);
   L.~G.~Almeida, S.~J.~Lee, G.~Perez, G.~Sterman and I.~Sung,
  arXiv:1006.2035;
  ATLAS Collaboration ATL-PHYS-PUB-2009-081;
  CMS Collaboration CMS-PAS-JME-09-001;
  K.~Rehermann and B.~Tweedie,
  arXiv:1007.222.
  
\bibitem{cmsnote}
M.~Aldaya et al., CMS NOTE 2006/106.


  \bibitem{Sjostrand:2006za}
  T.~Sjostrand, S.~Mrenna and P.~Z.~Skands,
  JHEP {\bf 0605}, 026 (2006).
  

\bibitem{higlu}
 M.~Spira,
  Nucl.\ Instrum.\ Meth.\  A {\bf 389}, 357 (1997).
  
\bibitem{wbfnlo}
    T.~Figy, C.~Oleari and D.~Zeppenfeld,
  Phys.\ Rev.\  D {\bf 68}, 073005 (2003)
  [arXiv:hep-ph/0306109].
   M.~Ciccolini, A.~Denner and S.~Dittmaier,
  Phys.\ Rev.\ Lett.\  {\bf 99}, 161803 (2007);
    B.~Jager, C.~Oleari and D.~Zeppenfeld,
  Phys.\ Rev.\  D {\bf 73}, 113006 (2006);
    K.~Arnold {\it et al.},
  Comput.\ Phys.\ Commun.\  {\bf 180}, 1661 (2009).
  
\bibitem{powheg}
  S.~Alioli, P.~Nason, C.~Oleari and E.~Re,
  JHEP {\bf 0904}, 002 (2009).

\bibitem{Alwall:2007st}
  J.~Alwall {\it et al.},
  JHEP {\bf 0709}, 028 (2007).

\bibitem{mcfm}
  J.~M.~Campbell and R.~K.~Ellis,
  Phys.\ Rev.\  D {\bf 60}, 113006 (1999);
    J.~M.~Campbell and R.~K.~Ellis,
  arXiv:1007.3492 [hep-ph];
http://mcfm.fnal.gov/.

\bibitem{Cacciari:2008zb}
  M.~Cacciari, S.~Frixione, M.~L.~Mangano, P.~Nason and G.~Ridolfi,
  JHEP {\bf 0809}, 127 (2008)
  [arXiv:0804.2800 [hep-ph]].

\bibitem{Dokshitzer:1997in}
  Y.~L.~Dokshitzer, G.~D.~Leder, S.~Moretti and B.~R.~Webber,
  JHEP {\bf 9708}, 001 (1997);
  M.~Wobisch and T.~Wengler,
  arXiv:hep-ph/9907280.


  \bibitem{Cacciari:2005hq}
  M.~Cacciari and G.~P.~Salam,
  Phys.\ Lett.\  B {\bf 641}, 57 (2006).
  M. Cacciari, G. P. Salam and G. Soyez, http://fastjet.fr
  
  
  \bibitem{antiKT}
  M.~Cacciari, G.~P.~Salam and G.~Soyez,
  JHEP {\bf 0804}, 063 (2008).
  
  \bibitem{KTjet}
  S.~D.~Ellis and D.~E.~Soper,
  Phys.\ Rev.\  D {\bf 48}, 3160 (1993).

\bibitem{Kribs:2007nz}
  G.~D.~Kribs, T.~Plehn, M.~Spannowsky and T.~M.~P.~Tait,
  Phys.\ Rev.\  D {\bf 76}, 075016 (2007).
 
 \bibitem{He:2001tp}
  H.~J.~He, N.~Polonsky and S.~f.~Su,
  Phys.\ Rev.\  D {\bf 64}, 053004 (2001);
    V.~A.~Novikov, L.~B.~Okun, A.~N.~Rozanov and M.~I.~Vysotsky,
  Phys.\ Lett.\  B {\bf 529}, 111 (2002);
    V.~A.~Novikov, L.~B.~Okun, A.~N.~Rozanov and M.~I.~Vysotsky,
  JETP Lett.\  {\bf 76}, 127 (2002)
  [Pisma Zh.\ Eksp.\ Teor.\ Fiz.\  {\bf 76}, 158 (2002)].

 \bibitem{4genmass}
http://www-cdf.fnal.gov/physics/new/top/2005/ljets/trpime/gen6/public.html;
  M.~Sher,
  Phys.\ Rev.\  D {\bf 61}, 057303 (2000);
  A.~Arhrib and W.~S.~Hou,
  Phys.\ Rev.\  D {\bf 64}, 073016 (2001);
  T.~Aaltonen {\it et al.}  [CDF Collaboration],
  Phys.\ Rev.\  D {\bf 76}, 072006 (2007).


\end{thebibliography}
\end{document}